\input{aipcheck.tex}

\documentclass[final]{aipproc}

\layoutstyle{6x9}

\usepackage{graphicx}

\newcommand{\apj}{ApJ}

\newcommand{\be}{\begin{equation}}
\newcommand{\ee}{\end{equation}}

\newcommand{\ba}{\begin{eqnarray}}
\newcommand{\ea}{\end{eqnarray}}
\newcommand{\om}{\omega}

\newcommand\etal{\textit{et al.\ }}
\newcommand\eg{\textit{e.g.\ }}

\begin{document}

\title{Electromagnetic (versus fireball) model of GRBs }

\author{M. LYUTIKOV}{
address={
Physics Department, McGill University, Montreal, QC, H3A 2T81 Canada}
}

\begin{abstract}
We briefly review the  electromagnetic model  of Gamma Ray Bursts  and then
discuss how various models account for high prompt polarization. 
We argue that if  polarization is  confirmed at a level $\Pi \geq 10\%$
the  internal shock model 
is excluded.
\end{abstract}

\date{\today}

\maketitle

\section{Electromagnetic model}

\begin{figure}
\includegraphics[width=0.6\linewidth]{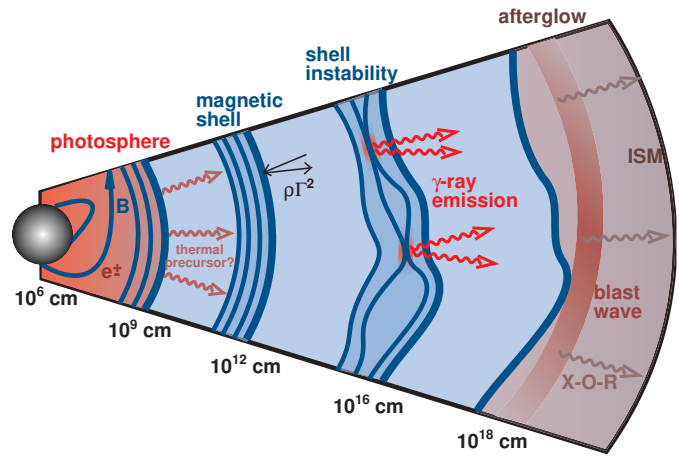}
\caption{Overview of the electromagnetic model}
\label{GRB-global}
\end{figure}

Electromagnetic model interprets 
Gamma Ray Bursts (GRBs) as 
 relativistic, electromagnetic explosions \cite{Cold}, 
(also Lyutikov \& Blandford, in preparation),  see Fig. \ref{GRB-global}.
It is  assumed that rotating, relativistic, stellar-mass
progenitor loses much of its rotational energy in the form of a
Poynting flux during the active period lasting $\sim 100$~sec.
The  energy to power the GRBs comes eventually from
 the rotational energy of the progenitor, converted  into
magnetic energy by the dynamo  action of the  unipolar inductor, so that
  the central  source acts as a power-supply generating
a current flow (along the axis, the surface of the  bubble
 and  the equator).
Initially non-spherically symmetric, 
electromagnetically dominated bubble expands non-relativistically inside the
star,  most rapidly  along the rotational axis of the progenitor. 
 The velocity of expansion of the bubble is determined by the pressure balance
on the contact between magnetic pressure in the bubble and
the ram pressure of the stellar material.
After the bubble  break out from the stellar surface
and most of 
the electron-positron pairs necessarily   present in the initial outflow
quickly annihilate the  bubble  expansion becomes  highly 
relativistic.
After the end of the  source activity most   of the magnetic energy
is concentrated in a thin shell inside the contact discontinuity between the
ejecta and the shocked circumstellar material. 
The 
 electromagnetic shell pushes ahead of it  a relativistic blast wave
into the circumstellar medium. 
Current-driven instabilities develop in this shell at a radius $\sim 3\times10^{16}$~cm 
and lead to acceleration of  pairs which are responsible for the $\gamma$-ray burst.
At larger radii the energy contained in the electromagnetic shell is mostly  transferred
to the preceding  blast wave.
Particles accelerated at the fluid shock may combine with electromagnetic 
field from the electromagnetic shell to produce the afterglow emission.

Electromagnetic model  produces ``structured jet'' 
with  energy $E_\Omega\propto \sin^{-2}\theta$  in a natural way
(in fact, there is no proper  ``jet'', but  non-spherical outflow and non-spherical
shock wave); there is no problem with ``orphan afterglow'' since 
GRBs are produced over large solid angle; X-ray flashes are interpreted 
as GRBs seen ``from the side'', but their total energetics should be comparable
to proper GRBs; 
  the model  can  qualitatively 
reproduce hard-to-soft  spectral evolution as a synchrotron  emission in 
ever decreasing magnetic field $B \propto \sqrt{L} /r $ ($L$ is luminosity, $r$ is emission
radius),
 akin  to "radius-to-frequency mapping" in radio
pulsars; similarly, the    correlation   $E_{peak} \sim \sqrt{L}$ 
is also a natural consequence. Finally, high polarization of prompt
emission  may also be  produced  \cite{lyu03c}
(it should  correlate with the spectral index; 
if there is a mixing between circumstellar material and ejecta, \eg due to Richtmyer-Meshkov
instability, 
and  if optical polarization is seen, then the  position angles  of the prompt emission
 and afterglow
 should  coincide and be
constant over time; fractional
polarization should 
be independent of the ``jet break'' time, 
but may show variations due to turbulent mixing).

\section{Polarization of prompt emission}

A very high linear polarization (nominally 80\%) has been reported in
RHESSI observations of GRB021206 \citep{coburn03}. 
Although the uncertainty in the measured
 polarization is large and  a   degree of reservation about the very result  is necessary,
the observation, {\bf if correct}, puts strong constraints
on the GRB models. In particular, it is  inconsistent with  the internal shock model, as
we argue below.

High polarization fraction cannot be naturally produced in a fireball model, since
magnetic fields expected in this model are produced on small microscopic scales.
In order to account for high polarization fraction  of the  prompt
 emission the 
fireball model needs to make  four  limiting assumptions three of which
are made specifically in order to explain polarization (some, but not all are listed
in \cite{Nakar}). 

First,
the turbulent magnetic fields, generated presumably by the Weibel instability, 
are  assumed to be exactly two-dimensional in the whole shocked region. 
Weibel instability \cite{med99}
indeed produces one-dimensional current filaments oriented normally to the
surface of the shock front but  the typical coherence size of magnetic field
is ion skin depth $\delta_i = c/\om_{p,i} = (m_i c^2/e) \Gamma  r \sqrt{c/L} 
\sim 5 \, {\rm cm} \, 
  L_{50}^{1/2} \Gamma_2 r_{12} $ (when the  shock scale is 
 $\sim r/\Gamma\sim 10^{10}$ cm,
nine orders larger).
Two dimensional inverse cascade that sets in after the field generation does increase field
  coherence  to tens or hundreds of ion skin depths,
but it  still remains microscopically small. Numerical
 simulations of the  Weibel turbulence  cannot run for long
enough times  to describe the late evolution of currents, still
we consider it unreasonable to expect that current will support alignment
on such  large scales. 
One of the reasons is that 
the postshock material is expected to be strongly  MHD  turbulent.
In fact, in  the fireball model MHD turbulence is {\it needed} 
 in order to accelerate particle by  the Fermi mechanism. This turbulence
will easily  randomize  subtle  current structures.
 Thus,  the  finely-aligned,   one-dimensional 
 currents and the presence of turbulence 
  necessary  to accelerate particles 
are in contradiction in the fireball model.
In addition, one may expect that oppositely directed
currents  created by the Weibel instability will eventually  close-up creating
three-dimensional magnetic
 structures.

Secondly, in order to produce high polarization in the fireball model
 the two-dimensional turbulent magnetic field should be viewed
 ''from the side'', so that
the  line of sight lies in the plane of the field. This imposes a constraint
on the position of the observer: the viewing angle with respect to the jet axis 
should be $ \theta_{ob} \sim 1/\Gamma$.
Thirdly, 
 in order to see  the emitting layer quasiplanar, the jet opening angle 
 should   be very small $\Delta \theta \leq 1/\Gamma$. 
Both these assumptions are not generic
 to the fireball model and are made exclusively in order to 
maximize polarization. 
(Generally,
if one needs to assume $\Delta \theta, \, \theta_{ob} \leq 1/\Gamma$
then the emission mechanism better be inverse Compton.)

Fourthly, since the observed burst was multi-peaked, many emitting shells
are required. In order to reproduce
high polarization 
  is assumed that all shells move with the same Lorentz factor \cite{Nakar}. 
This assumption
runs contrary to the very basic 
postulate  of the fireball model that the emitting shells
are due to collision of material moving with  {\it different Lorentz factor},
 so that the
velocity of the resulting shocks must be different.

Thus, the fireball model needs to fine-tune  several parameters to   explain polarization.
An argument of exclusivity has  been invoked: the burst was not like any other burst,
 so it tells
nothing about the other bursts. 
This is virtually   equivalent to neglecting the results altogether.

Though one can possibly argue in favor of  chance coincidence of $\Delta \theta $ 
and $ \theta_{ob}$ 
(but the GRB  rate also goes up by $\Gamma^2 \sim 10^{4}$)
 one cannot bypass the problem with the spread in Lorentz factors
and turbulent randomization of fields.
In order to maximize polarization the 
spread in the Lorentz factors of the emitting shells
must be small. This, in principle, can be achieved by a carefully arrangement of shells
so that  collisions occur only 
between the two blobs that are moving with the same
Lorentz factors (if the source emits  shells with
 $\Gamma_1, \Gamma_2, \Gamma_1$ etc).   This is an extremely contrived situation. 
A more generic case is that the blobs' Lorentz factors are randomly distributed. 
In this case the fireball  model needs to thread a  thin line: larger polarization would
require smaller spread in Lorentz factors, but  then  the  energy 
available for dissipation is small, so that the total energy of the burst 
will be very large (and the burst GRB 021206 was unusually  luminous to start with), 
aggravating even further the efficiency problem of internal shocks.

In addition to the problems specific to the fireball model, 
there is a  kinematic depolarization of synchrotron radiation
due to inhomogeneous expansion velocity \cite{lyu03c} (it was taken into account by 
\cite{Nakar}).
Electrical vectors
of waves emitted by different parts of the flow are rotated by different amount during a boost
into observer's frame, so that the 
 {\it observed electric fields are generally  not orthogonal to 
the observed magnetic field}. 
Averaging over emitting volume reduces total polarization.

Thus, even if  turbulence downstream and current closure are completely neglected, 
the effects of relativistic kinematics, randomness of magnetic field, 
spread in Lorentz factors and (presumably) not a perfect fine-tuning of $\Delta \theta$ 
and $\theta_{ob}$ will all contribute to reduction in polarization. 
Generically, each of these effects will contribute a factor of two, so  that 
the resulting polarization will not exceed $\sim 10\%$. (Mathematically, when all
depolarization effects are minimized at once, polarization may be somewhat higher, but still
 $\leq 20\%$). 

\begin{figure}
\includegraphics[width=0.45\linewidth]{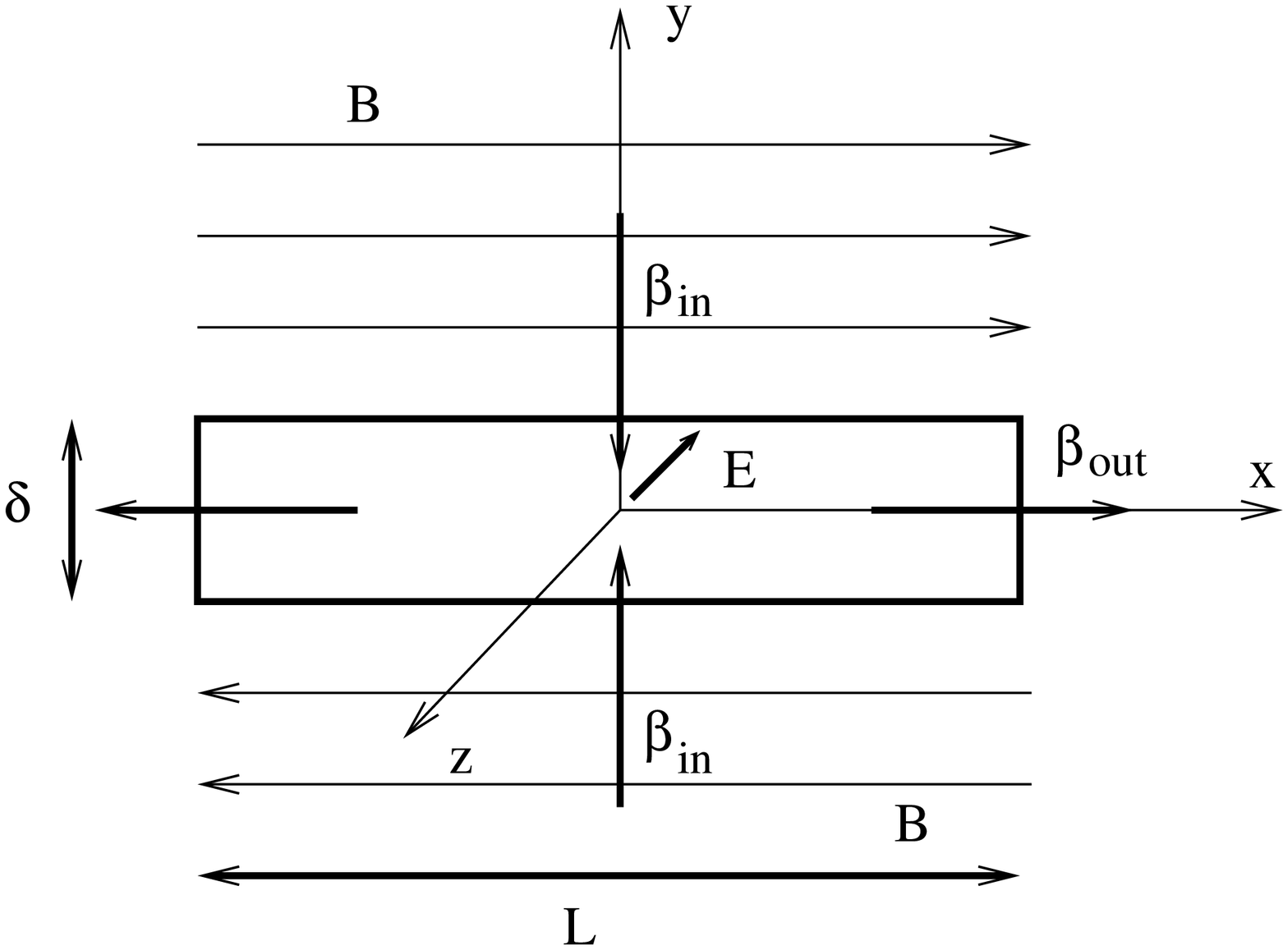}
\includegraphics[width=0.35\linewidth]{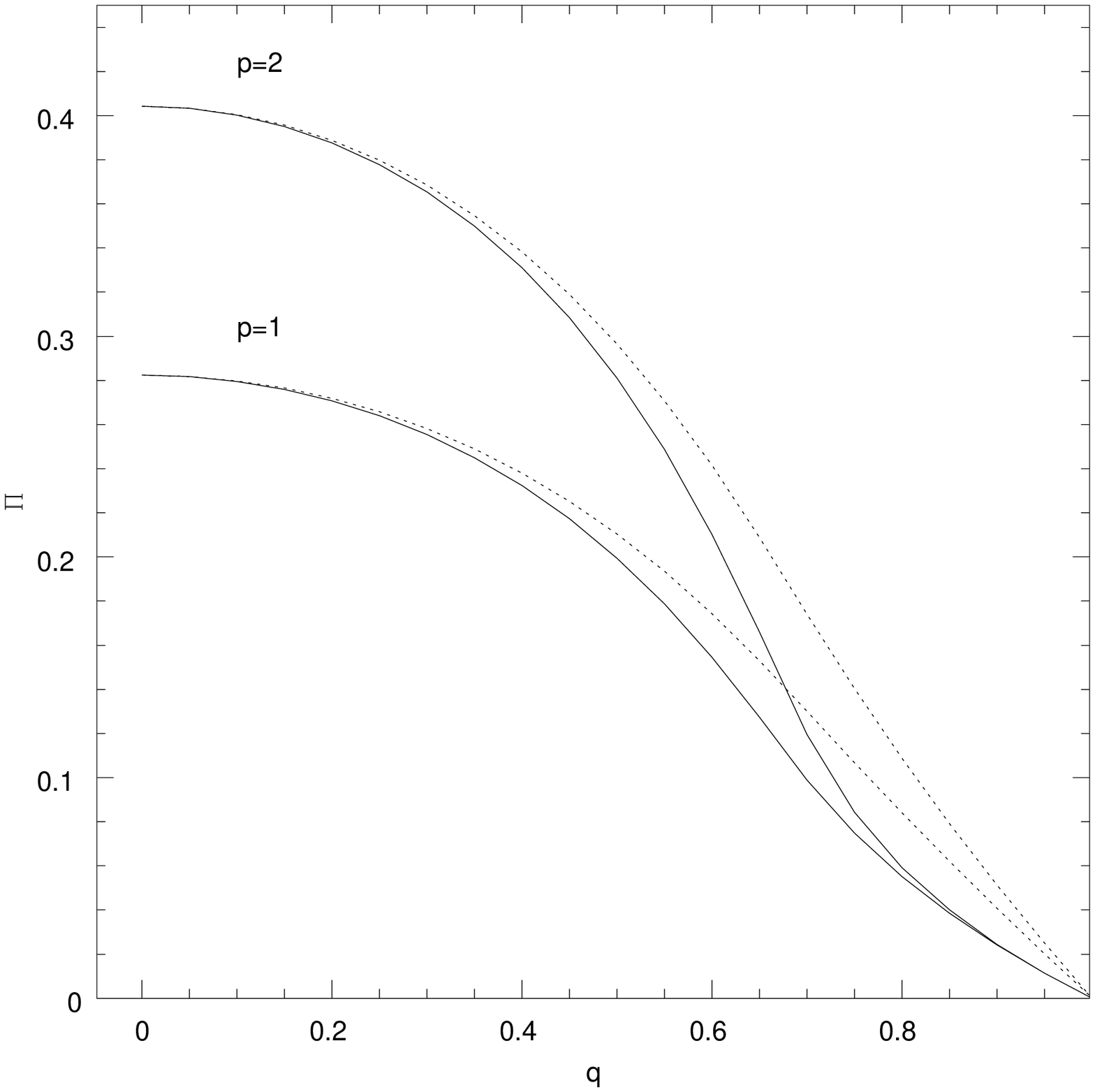}
  \caption{(a) An example of ordered inhomogeneous magnetic field (relativistic
Sweet-Parker model of reconnection, after \cite{lu03}).
 (b) Polarization fraction in the   electromagnetic model as a function of 
$q$, the ratio of the rms fluctuations to the total field
$q= \sqrt{ < B_{rn}^2>} /B $ for different values
of the particle power-law index $p$.
 Solid lines are for two-dimensional random magnetic field
confined to the  ${\bf e}_\theta- {\bf e}_\phi$ plane,
dashed lines  - for the three-dimensional random magnetic field.
 The large scale magnetic field is  $B_\phi$. For details see
\cite{lyu03c}. }
\label{f1}
\end{figure}

In  the case of electromagnetic model the  magnetic field
has a large coherence  scale, $\sim r$, so that within a visible patch
of linear size $r/\Gamma$ the field is quasihomogeneous.
There are also  
  depolarization effect. First, there is   kinematic depolarization  discussed 
above \cite{lyu03c}.
   Secondly,    possible  presence of random component
of magnetic field would lead to further decrease of polarization.
But generically  {\it random component 
is not needed in the electromagnetic models}. What is needed is  presence
of currents, so that the magnetic field is inhomogeneous, but it still
can be ordered. The field structure  of the Sweet-Parker reconnection layer
gives an excellent example (see Fig \ref{f1}.a).
Random component of the magnetic field is naturally expected and one should account for it. 
In fact, the very amount of the dissipated magnetic energy may be 
related to the random component of the field (\eg MacFadyen, these proceedings). 
 For efficient accelerations of electrons one  then would need 
 $\delta B /B  \sim 1$.
 In this case 
the total polarization decreases  (Fig. \ref{f1}.b),  remaining  reasonably
 high for  $\delta B /B \leq 1$.

\begin{figure}
\includegraphics[width=0.55\linewidth]{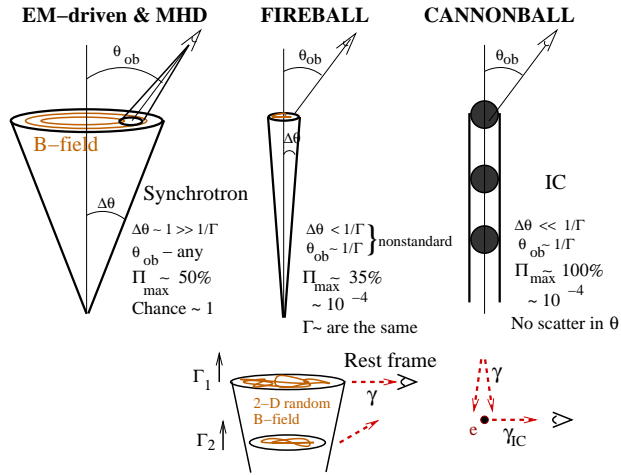}
\caption{Comparison of different models for prompt
 polarization. For the fireball and
cannonball models  the maximum polarization given  is for  a single emitting shell 
(or ball).}
\label{polarizAll-GRB}
\end{figure}

The three competing  models of GRBs are compared in Fig \ref{polarizAll-GRB}. 
Electromagnetic model is the best contender. It does not require fine tuning
of parameters, 
all observers should see large polarization regardless of the viewing angle,
Lorentz factor etc. The  only constraint
is that the random component of the  field  is not dominant.

The cannonball model \cite{dar03}
 (and other models invoking inverse Compton scattering \cite{eich03,laz03})
 can in principle produce very high polarization.
It does make  an unphysical (in our opinion)
assumptions $\Delta \theta, \, \theta_{ob} \leq 1/\Gamma$, but this was
 inherent to the model before polarization results came out.
There is a large degree of fine-tuning: it is assumed that all ''cannonballs''
are flying within an extremely narrow cone. If their directions were to have a
scatter $\Delta \theta  >  1/\Gamma \sim  10^{-3}$ polarization would  drop to zero.
Still, it has an advantage that in the best case it can produce up to 100\%
polarization. 

Internal shocks model is the weakest player in the field. 
It needs to make very contrived  and contradictory assumptions
to reproduce observations. 

\subsection{Is RHESSI polarization real?}
 
High polarization of prompt emission, if confirmed, would put  strong
constraint on the emission mechanisms and GRB models.
 Only inverse Compton may
produce polarization as high as 80\%. Electromagnetic models can
get to $\sim 50\%$, but with a  random component a comfortable range
is $\leq 30-40\%$. For internal shock  anything above  $\sim 10\%$ is unreasonable.
Obviously, the RHESSI polarization result is highly doubtful, so that an independent
confirmation or refutal is a must.  If the results are not confirmed, 
i.e. polarization is $\Pi \leq 20\%$, that won't exclude any model since a multitude
of depolarization effects may intervene.

I'd like to thank  R. Blandford, M. Medvedev, V. Pariev and D. Lazzati.

\end{document}